\documentclass[journal=jpcl,manuscript=letter]{achemso}
\usepackage[version=3]{mhchem}
\usepackage{caption}
\usepackage{subcaption}
\usepackage{hyperref}
\usepackage{natbib}
\usepackage{siunitx}
\usepackage{tabularx}
\usepackage{amsmath}
\usepackage{amssymb}
\usepackage{xcolor}
\usepackage{threeparttable}
\usepackage{multirow}
\usepackage{booktabs}
\usepackage{subcaption}
\usepackage{braket}
\usepackage{derivative}
\usepackage{physics}
\usepackage{gensymb}

\author{Arpan Kundu}
\affiliation{Pritzker School of Molecular Engineering, The University of Chicago, Chicago, Illinois 60637, United States}
\altaffiliation{These two authors contributed equally.}
\email{arpan.kundu@gmail.com}
\author{Francesco Martinelli}
\affiliation{Pritzker School of Molecular Engineering, The University of Chicago, Chicago, Illinois 60637, United States}
\alsoaffiliation{Present Address: Materials Theory, Department of Materials, ETH Z\"{u}rich, Wolfgang-Pauli-Strasse 27, 8093 Z\"{u}rich, Switzerland}
\altaffiliation{These two authors contributed equally.}
\author{Giulia Galli}
\affiliation{Pritzker School of Molecular Engineering, The University of Chicago, Chicago, Illinois 60637, United States}
\alsoaffiliation{Department of Chemistry, University of Chicago, Chicago, Illinois 60637, United States}
\alsoaffiliation{Materials Science Division and Center for Molecular Engineering, Argonne National Laboratory, Lemont, Illinois 60439, United States}
\email{gagalli@uchicago.edu}

\title{ Designing Optically Addressable Nitrogen-Vacancy Centers in Ultra-Small Nanodiamonds: Insights from First-Principles Calculations}
\date{\today}
\begin{document}
\maketitle

\begin{abstract}
Ultrasmall nanodiamonds (USNDs) are promising platforms for fluorescent and quantum sensing applications. Here we present first-principles electronic structure calculations of color centers in USNDs, specifically the nitrogen-vacancy (NV-)  and we investigate their optical addressability as a function of the surface termination. We consider both isolated nanoparticles and arrays of USNDs with different degrees of packing, and we include quantum vibronic effects in our analysis, using stochastic methods. We find that the NV in USNDs can be stabilized in a negative charge state if the nanoparticles are terminated by fluorine, hydroxyl, and ether. While fluorine terminations can be used for fluorescent bio-tags, we suggest that hydroxyl and ether terminations are beneficial for quantum sensing applications. We also find that the NV- can be stabilized in arrays of USNDs when inter-particle separations are larger than the diameter of the nanoparticle. Interestingly, the phonon renormalizations of single-particle energy levels found in arrays contribute to the charge stability of negatively charged NV centers.
\end{abstract}

Fluorescent nanodiamonds\cite{Mochalin_Nat_Nanotech_2012,  Hsiao_Acc_Chem_Res_2016, Alkahtani_Nanophotonics_2018, Mochalin_book_2020} are interesting platforms for nanotechnology and emerging quantum technologies, as they show promising properties for a wide range of applications, including cell tracking,\cite{Fu_PNAS_2007, Falkaris_ACS_Nano_2009, Wu_Nat_Nano_2013} super-resolution imaging,\cite{Rittweger_Nat_Photonics_2009, Han_Nano_Lett_2009, Tzeng_Angew_2011}  long-term information storage,\cite{Laube_ACS_Nano_2023} and nanoscale quantum sensing\cite{Schirhagl_2014, Barry_Rev_Mod_Phys_2020, Wu_Adv_Sci_2022, Aslam_Nat_Rev_Phys_2023, Rovny2024}.  Ultra-small nanodiamonds (USNDs), with diameters below 5 nm, offer several, specific advantages.\cite{Chang_ACS_Nano_2022}. For example, if color centers can be stabilized in USNDs,  they always reside near the surface, thus being more sensitive to environmental changes and enhancing the effectiveness of the nanoparticles as quantum sensors.\cite{Rodgers_MRL_Bull_2021} Additionally, the small size of USNDs allows for higher penetration rates through narrow pores in biological systems, facilitating numerous biomedical applications.\cite{Turcheniuk_Nanotechnology_2017} 

Among color centers in bulk diamond, the nitrogen-vacancy (NV) centers, which may be stabilized in neutral (NV$^0$), negatively (NV$^-$), or positively  (NV$^+$) charged states, are the most well-studied and the NV$^-$ has found numerous applications.  When excited with a green laser (550 nm), the NV$^-$ center undergoes a $^3A_2 \rightarrow{}^3E$ electronic transition, resulting in a red fluorescence peak at 685 nm with a strong quantum yield and a long fluorescence lifetime (approximately 20 ns), facilitating its utilization as a biomarker\cite{Fu_PNAS_2007, Barnard_Analyst_2009, Falkaris_ACS_Nano_2009, Wu_Nat_Nano_2013} or in super-resolution imaging.\cite{Rittweger_Nat_Photonics_2009, Han_Nano_Lett_2009, Tzeng_Angew_2011} Additionally, the ability to read out the spin states ($m_s = 0$ and $\pm 1$) using optically detected magnetic resonance (ODMR) techniques, combined with a long spin-coherence time (up to milli-seconds), has led to the utilization of the NV$^-$ center as a quantum sensor \cite{Schirhagl_2014, Barry_Rev_Mod_Phys_2020, Wu_Adv_Sci_2022, Aslam_Nat_Rev_Phys_2023, commun, computing1, computing2, Rembold_AVS_Quantum_Sci_2020}. The NV$^0$ lacks many of the desirable properties of the NV$^-$ and the $\text{NV}^{-}\rightarrow\text{NV}^{0}$ charge conversion leads to fluorescence quenching and to a significantly decreased spin-coherence time,\cite{Grotz_Nat_Comm_2012, Okai_PRB_2012, Schreyvogel_Sci_Rep_2015, Bluvstein_PRL_2019, Yuan_PRR_2020, Garcia_ACS_Photonics_2020} affecting the optical addressability of the NV$^-$. Although the NV$^-$ center is stable in  bulk diamond and in nanodiamonds with diameter $>10$ nm, unfortunately, it is converted to its neutral state upon optical excitation if it resides close ($<10$ nm) to the surface\cite{Bradac_Nat_Nanotechnol_2010, Barnard_Mat_horizon_2014, Yuan_PRR_2020}.

Proposed strategies\cite{Kaviani_Nano_Lett_2014, Chou_Nano_Lett_2017, Kageura_Carbon_2022, Haruyama_APL_2023, Neethirajan_Nano_Lett_2023, Giri_ACS_2023, Geng_NPJ_2023, Kumar_ACS_Photonics_2024, Gorrini_Biosensors_2023} to stabilize  NV$^-$-centers close to a diamond surface include doping and surface modification through chemical functionalization, adsorption of molecules, and surface coating. So far, most studies focused on surfaces terminating bulk diamonds, while strategies for stabilizing NV$^-$ centers within USNDs remain unexplored.  Additionally, how the aggregation of USNDs into arrays impacts the properties of color centers is not well understood and it is an important question, in view of the interest in nanodiamond arrays for long-term optical storage\cite{Laube_ACS_Nano_2023} or quantum sensing of stress fields\cite{Ovartchaiyapong_Nat_Commun_2014, Teissier_PRL_2014, Barfuss_PRB_2019}. 

In this work, using density functional theory (DFT), we performed first-principles electronic structure calculations of NV$^-$ centers in USNDs to investigate their optical addressability as a function of the surface termination of the isolated nanoparticle. We also studied arrays of USNDs with different degrees of packing, and we investigated quantum vibronic effects, which are known to significantly alter the electronic structure of crystalline and amorphous diamond\cite{Kundu_PRM_2021, Han_JCTC_2021, Kundu_PNAS_2022}, diamondoids \cite{Kundu_JCTC_2023}, NV-centers in crystalline diamond.\cite{Han_JCTC_2022, Kundu_NV_JPCL_2024}.

\begin{figure}[hp]
\centering
\includegraphics[width=16.7cm]{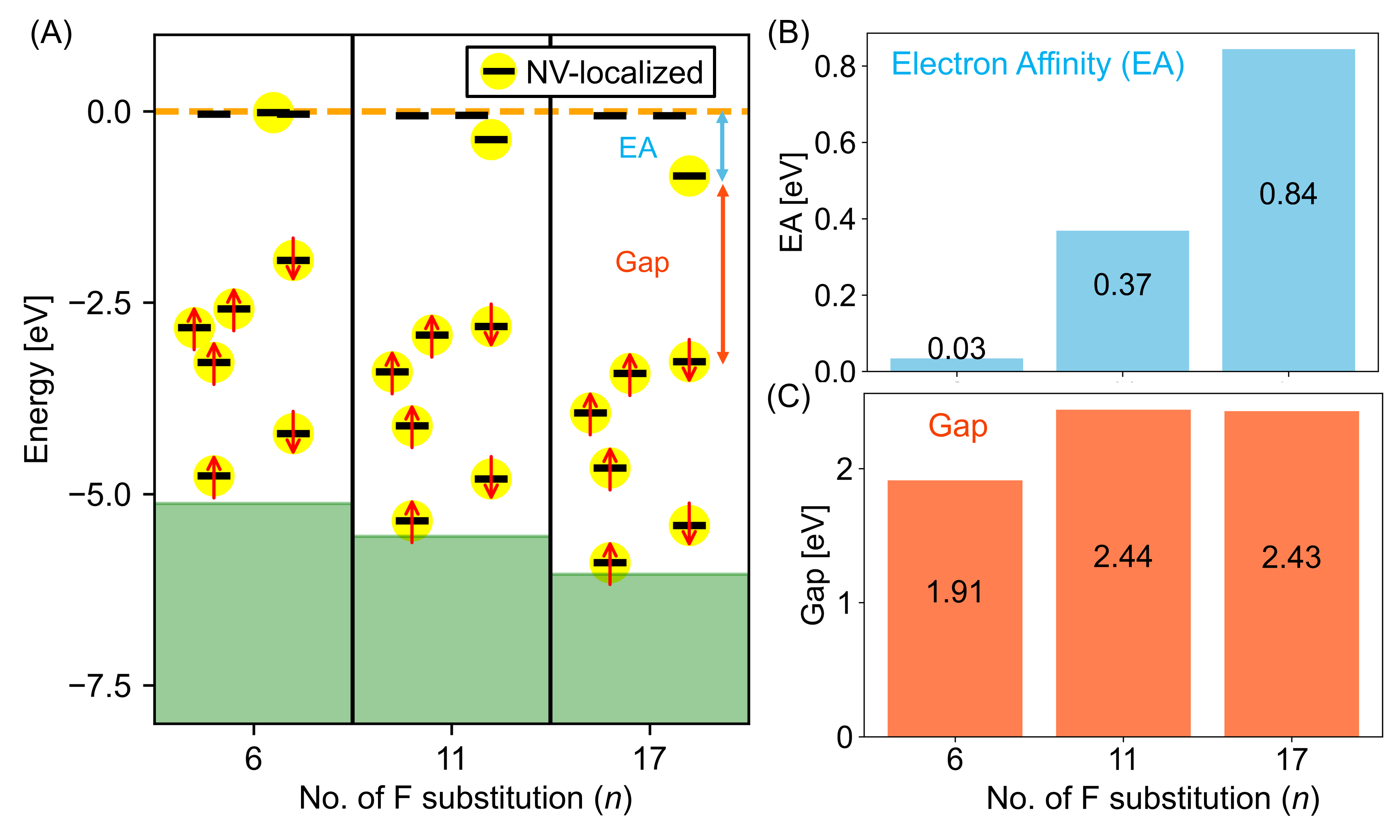}
\caption{Comparison of the electronic properties (obtained with the B3LYP hybrid functional) of various F-substituted isolated nanodiamond models (C\textsubscript{67}H\textsubscript{62-$n$}F\textsubscript{$n$}N)
with varying numbers ($n$) of F-substitutions: (A) single-particle energy levels (aligned to vacuum), (B) electron affinity, (C) HOMO-LUMO gap in the minority ($\beta$) spin channel. 
In panel A, NV-localized single-particle levels (with a localization factor larger than 0.5) are highlighted in yellow.  We computed the localization factor, \cite{Nan_JCTC_2022} $L=\int_{r \leq 1.5 , \text{\AA}} |\psi|^2 d^3r$ of each level by integrating the square of the Kohn-Sham wave function $(\psi)$ within a sphere of radius 1.5 \text{\AA} centered at the vacancy. }
\label{fig:F_term_B3LYP}
\end{figure}

{\it Isolated Nanodiamonds} We first investigated two clusters (C\textsubscript{33}H\textsubscript{36}N (0.7 nm) and C\textsubscript{67}H\textsubscript{62}N) ($\simeq$ 1 nm) with hydrogen-terminated surfaces, and we found that they both exhibit negative electron affinity (EA). This result is not surprising, since H-terminated diamond surfaces also have negative EA,\cite{Maier_PRB_2001} which can be turned into positive by, for example, fluorine substitutions\cite{Rietwyk_APL_2013},   ultimately leading to a significant increase of the NV$^-$/NV$^0$ ratio.\cite{Cui_APL_2013}  It is then natural to consider electron-withdrawing groups (EWGs) instead of hydrogen, to obtain nanoparticles with positive electron affinity.  Hence we turned to substituting some hydrogen atoms of the C\textsubscript{67}H\textsubscript{62}N nanodiamond model with an electronegative element: fluorine. We gradually increased the degree of substitution to 100\%, optimized the geometries using the Perdew-Burke-Ernzerhof (PBE) functional,\cite{PBE_Perdew_PRL_1996_1, PBE_Perdew_PRL_1996_2} and subsequently, computed the electronic properties of the nanoparticle with both PBE and with hybrid functionals. In Fig. \ref{fig:F_term_B3LYP} we compare (A) single-particle energy levels, (B) electron affinity (defined as the energy difference between the vacuum level of the nanoparticle and the lowest unoccupied molecular orbitals (LUMO) in the minority ($\beta$) spin channel), and (C) HOMO-LUMO gaps (defined as the energy difference between highest occupied molecular orbital (HOMO) of the defect and LUMO of the defect in the $\beta$-spin channel) obtained with the B3LYP\cite{B3LYP_PRA_1988, B3LYP_PRB_1988, B3LYP_JCP_1993} hybrid functional.

Fig. \ref{fig:F_term_B3LYP} shows that as the number of F atoms at the surface increases, the EA of the NV-localized LUMO and the HOMO-LUMO gap increase.  While a sufficiently positive EA is essential for the charge stability of the NV$^-$-center, an optimally tuned HOMO-LUMO gap is also required. Our results suggest that charge stability and HOMO-LUMO gap tuning are possible for shallow NV$^-$-centers in nanodiamonds by controlling the concentration of fluorine at the surface. Interestingly, earlier calculations on F-terminated surfaces of crystalline diamond showed the presence of surface-related localized acceptor states within the band gap, possibly due to the chosen geometrical configurations at the surface; surface acceptors are undesirable as they limit sensing applications due to the bleaching or blinking of NV$^-$ fluorescence signals.\cite{Kaviani_Nano_Lett_2014} The absence of such states in our calculations suggests that hosting an NV$^-$-center within a spherical F-terminated small nanodiamond may be a better alternative than hosting it in bulk diamond with F-terminated surfaces.  

While F-substitutions may be appropriate for fluorescence sensing applications, they are not desirable for quantum information science applications. The only available non-radioactive isotope of fluorine (\textsuperscript{19}F) has a half-integral nuclear spin, leading to a short coherence time of the electronic spin residing at the NV\textsuperscript{-}-center. Therefore, surface terminations with other EWGs containing elements that have at least one non-radioactive isotope with an integral nuclear spin (\textit{e.g.}, sulfur and oxygen, etc.) are preferred.

Hence we considered three additional functional groups (--X) terminating the 1 nm nanodiamond model, with the chemical formula C\textsubscript{67}H\textsubscript{45}X\textsubscript{17}N. Each functional group (--X) belongs to a distinct class, and it exerts different electronic effects on the nanodiamond: 
(i) Hydroxyl (--OH) group: it has weaker electron-withdrawing  inductive ($-I$) effects than --F. In addition, it has also an electron-donating resonance (+R) effect. 
(ii) Aldehyde (--CHO) group: it has strong electron-withdrawing inductive ($-I$) and resonance ($-R$) effects.
(iii) Thiol (--SH) group: it has a much weaker $-I$ effect than --OH groups, as  S  is less electronegative than O. The $+R$ effects of --SH groups is also weaker because the lone-pairs on the S sites are located at  3sp\textsuperscript{3} hybrid orbitals, with limited overlap with the 2sp\textsuperscript{n} ($n=2,3$)  hybrid orbitals of carbon.
  
We note that thiols, being soft bases, can form robust bonds with noble metal surfaces. This principle can be utilized to synthesize self-assembled monolayers, which are promising for molecular electronics\cite{Tour_Acc_Chem_Res_2000, Villan_Chem_Rev_2017}  and qubit \cite{Tesi_Adv_Mat_2023} applications,  and such synthetic route has spurred a growing interest in functionalizing nanodiamonds with thiols.\cite{Tkchenko_Org_Lett_2006, Hsu_ACS_appl_mat_interfaces_2014, Parker_Diam_Relat_Mater_2021}  

\begin{figure}[htbp]
\centering
\includegraphics[width=14cm]{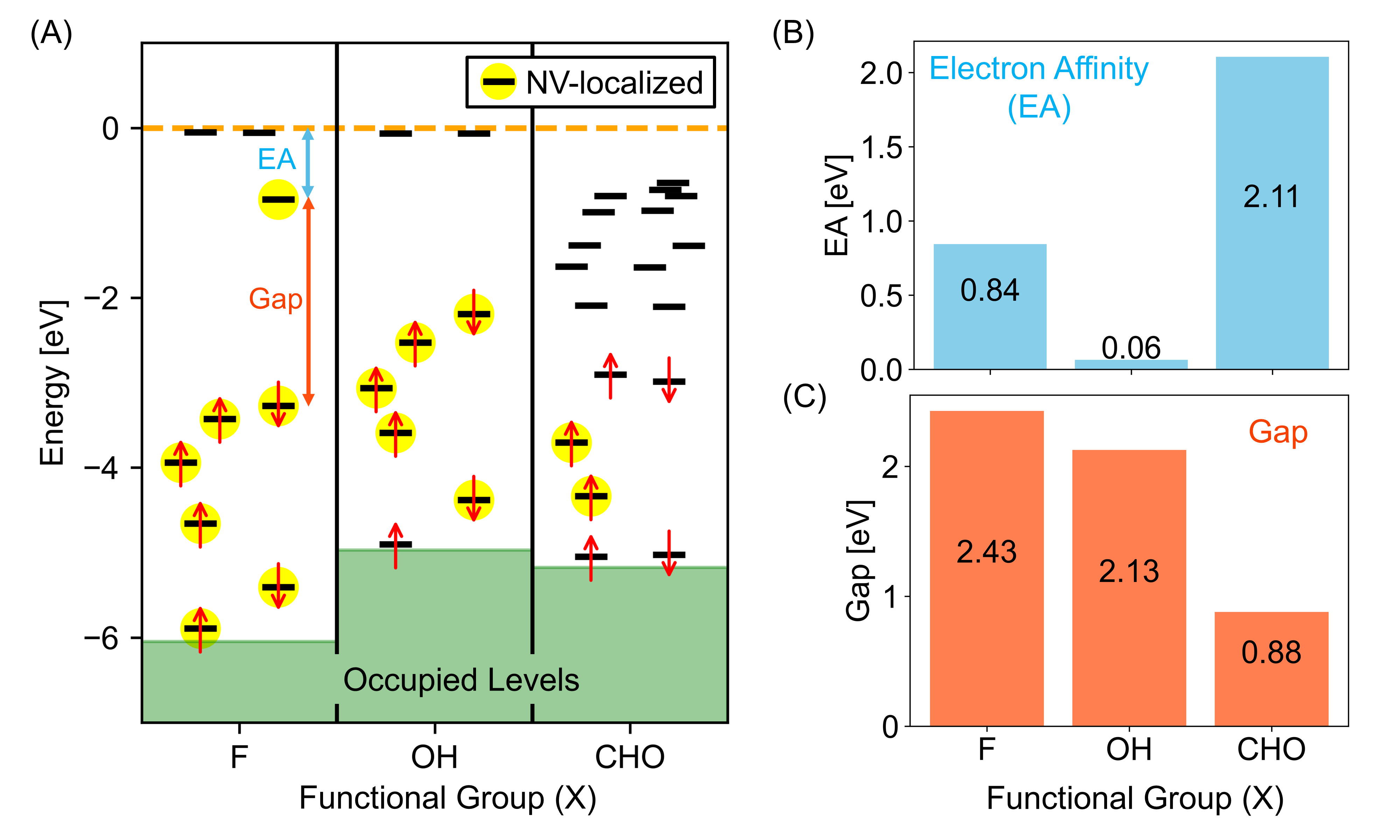}
\caption{Comparison of the electronic properties (computed with the B3LYP hybrid functional) of  isolated nanodiamond models (C\textsubscript{67}H\textsubscript{45}X\textsubscript{17}N)
obtained by substituting different functional groups (X) at the surface: (A) single-particle energy levels (aligned to vacuum), (B) electron affinity, (C) HOMO-LUMO gap in the minority ($\beta$) spin channel. Localization of states is defined as in Fig. \ref{fig:F_term_B3LYP}.}
\label{fig:all_term_B3LYP}
\end{figure}

We used the same topological positions as F-terminations and optimized the geometry of the USNDs using the PBE functional and subsequently, determined the electronic structure using PBE and hybrid functionals. Fig. \ref{fig:all_term_B3LYP} shows the (A) single-particle energy levels, (B) electron affinity, and (C) HOMO-LUMO gap ($\beta$-spin channel) for the terminations listed above, as obtained with the B3LYP hybrid functional.

Fig. \ref{fig:all_term_B3LYP} shows that --OH terminations do not stabilize NV-localized unoccupied levels in the $\beta$-spin channel and the EA of the nanoparticle is almost close to zero. As a result, the NV$^-$-center would not be optically addressable. Despite substantially raising the EA, the  --CHO group affects the localization (at the NV-center) of the occupied (unoccupied) levels near the HOMO(LUMO). Visualization of these levels reveals that they are essentially $\pi^*$ orbitals associated with the  --CHO terminating groups.
For the --SH terminated nanodiamonds,  we were unable to converge the electronic structure calculations using hybrid density functionals. Nevertheless, the results obtained with the PBE functional show that despite achieving a more positive EA than with -OH terminations, the --SH groups lead to HOMO-LUMO orbitals that are not localized at the NV-center due to the hybridization of the $3s3p$(S) levels. This lack of localization leads to an undesirable charge transfer from the NV-center to the surface. Interestingly, in previous work, the presence of acceptor surface states has been attributed to the blinking and bleaching of shallow  NV$^-$-center beneath the surfaces of crystalline diamonds.\cite{Kaviani_Nano_Lett_2014}  Therefore, our findings suggest that, in the case of USNDs, the --CHO and  --SH terminations may not be useful for either fluorescence sensing or quantum information science applications.

\begin{figure}[htbp]
\centering
\includegraphics[width=16cm]{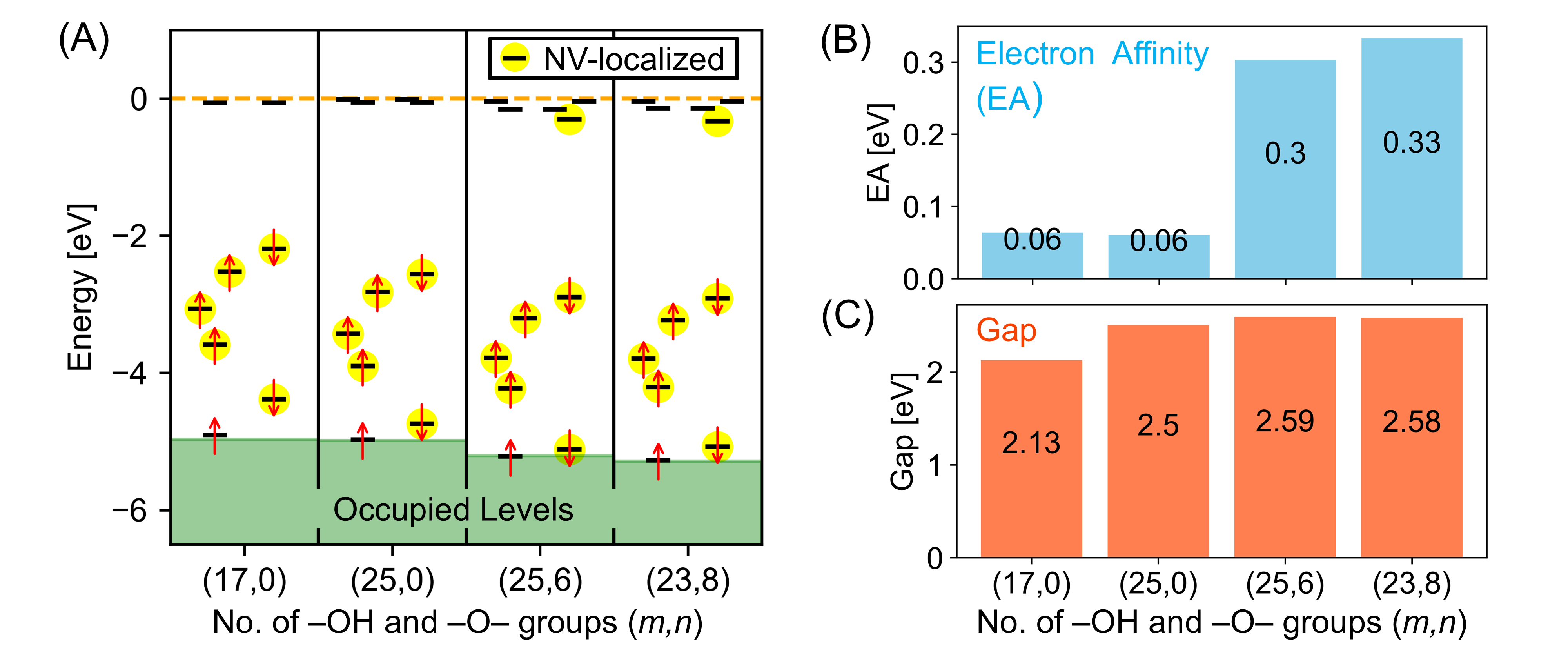}
\caption{Comparison of the electronic properties (obtained with the B3LYP hybrid functional) of  isolated nanodiamond models with chemical formula C\textsubscript{$67-n$}H\textsubscript{$62-m-2n$}(OH)\textsubscript{m}O\textsubscript{$n$}N
having different numbers of hydroxyl (OH) and ether (O) functional groups : (A) single-particle energy levels (aligned to vacuum), (B) electron affinity, (C) HOMO-LUMO gap in the minority ($\beta$) spin channel. Localization of states is defined as in Fig. \ref{fig:F_term_B3LYP}.}
\label{fig:OHO_term_B3LYP}
\end{figure}

Previous first-principle calculations suggested that a combination of hydroxyl and ether surface terminations for the near-surface NV$^-$-center hosted in bulk diamond is optimal for sensing applications.\cite{Kaviani_Nano_Lett_2014} Further, recent X-ray photoelectron spectroscopy (XPS) data suggests the presence of ether groups (C--O--C) in hydroxyl-terminated diamond surfaces prepared using high-voltage hydroxide ion treatments.\cite{Li_Appl_Surf_Sci_2023} Therefore, in addition to the hydroxyl groups, we considered introducing ether (--O--) groups to the NV$^-$-nanodiamond surfaces to stabilize NV-localized unoccupied levels below the vacuum level. Fig. \ref{fig:OHO_term_B3LYP} highlights the B3LYP results for the resulting nanodiamond clusters with the chemical formula: C\textsubscript{$67-n$}H\textsubscript{$62-m-2n$}(OH)\textsubscript{m}O\textsubscript{$n$}N, where $m$ and $n$ represent the number of hydroxyl and ether groups, respectively.

After introducing at least 23 --OH groups and 6--8  --O-- groups at the surface of the nanocluster, we find that NV- levels are localized and the USND has positive EA. Notably, after introducing 6 --O-- groups at the nanodiamond surface already containing 25 --OH groups, the gap increases only marginally (by 0.09 eV), but we could identify NV-localized orbitals  0.3 eV below the vacuum level. These results found at the B3LYP level are also confirmed when using other functionals, including PBE and PBE0.
These findings demonstrate that similar to near-surface  NV$^-$-center hosted in crystalline diamond,\cite{Kaviani_Nano_Lett_2014} a combination of --OH and --O-- surface terminations can stabilize NV-localized unoccupied levels, opening the possibility of engineering optically addressable NV$^-$-centers within small nanodiamond clusters.   

\begin{figure}[htbp]
\centering
\includegraphics[width=8.3cm]{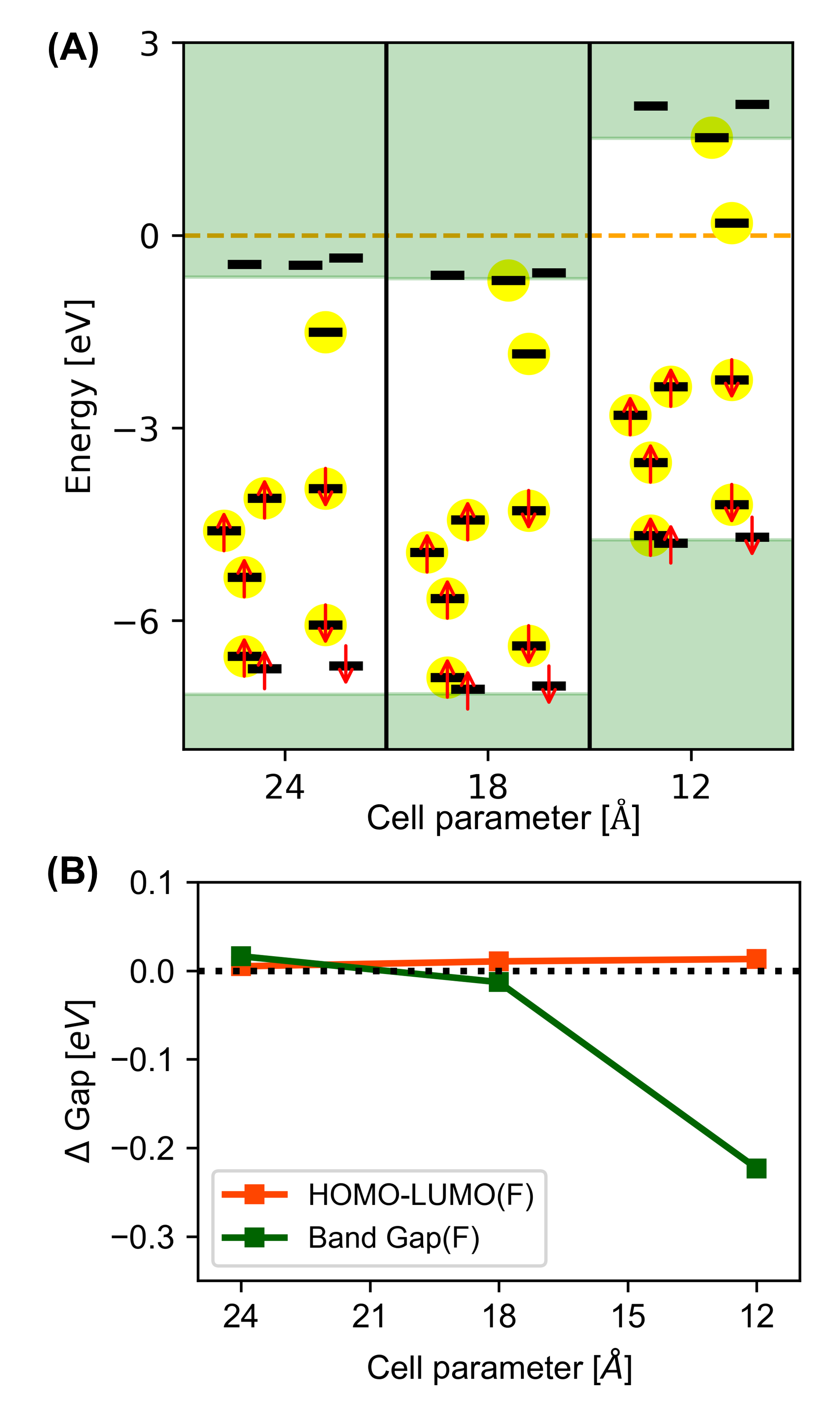}
\caption{ Electronic properties of F-terminated nanodiamond  (C\textsubscript{67}H\textsubscript{45}F\textsubscript{17}N) arrays hosting NV$^{-}$ -centers as obtained with B3LYP functional (A) Single particle energy levels as a function of decreasing cell parameters. The green-shaded regions are the position of the valence and conduction bands of the respective pristine  (\textit{ i.e.} without the NV$^{-}$-center) nanodiamond array.s The zero energy reference is defined as the highest electrostatic potential within the unit cell, computed along the lattice direction where this potential achieves the lowest maximum value compared to other directions. Localization of states is defined as in Fig. \ref{fig:F_term_B3LYP}. (B) The change in band gaps (for the pristine nanodiamond arrays) and HOMO-LUMO gaps in the $\beta$-spin channel (for the nanodiamond arrays with an NV$^{-}$) with respect to the isolated nanodiamonds are plotted as a function of decreasing cell parameters.  }
\label{fig:crystal_B3LYP}
\end{figure}

{\it Arrays of nanodiamonds} -- We now turn to discuss USNDs assembled in arrays. We considered representative nanoparticles with positive electron affinity ( F-terminated nanodiamonds: C\textsubscript{67}H\textsubscript{45}F\textsubscript{17}N); we arranged them into a  body-centered cubic (BCC) lattice (chosen for computational simplicity), and we considered the primitive unit cell of such a lattice. We studied the electronic structure of the lattice as a function of the size of the primitive unit cell, varying it from 24 {\AA} to 12 {\AA}, to understand whether an NV- center may be stabilized on a given site (nanoparticle) or whether charge hopping between nanoparticles is likely to occur. As for isolated nanoparticles, after optimizing the geometry with the PBE functional we computed the electronic structure with hybrid functionals. Fig. \ref{fig:crystal_B3LYP} shows the (A) single-particle energy levels and (B) HOMO-LOMO ($\beta$-spin channel) and band gaps for the F-terminated crystals using the B3LYP functional. While we discuss only the B3LYP results below, our findings are the same irrespective of the functionals used.

 In Fig. \ref{fig:crystal_B3LYP}A, the value of the maximum potential within the unit cell, which defines the potential energy barrier for an electron to be transferred from one nanocluster to another, is subtracted from each energy level computed for the crystals.   
As the USNDs approach to form a condensed phase, we observe a steady decrease in the energy of NV-localized unoccupied orbitals, which lie below the conduction band of the host nanocrystal. This energy decrease continues until a cell length of 15 {\AA}. 

Our findings show that for packing distances larger than the cluster's diameter, the NV- remains stable in arrays. However, when cell lengths become comparable to the diameter of the nanoparticles, the energy barrier for charge (spin) hopping between adjacent nanoclusters decreases, resulting in a higher probability of such hopping. Consequently, the charge (spin) would not remain localized into a single nanocluster, leading to challenges in the spin-readout process. However, increased charge (spin) hopping could be harnessed for charge (spin) transport applications, such as developing an optical spin-valve. \cite{optical_spin_valve}   

\begin{figure}[h]
\centering
\includegraphics[width=8.3cm]{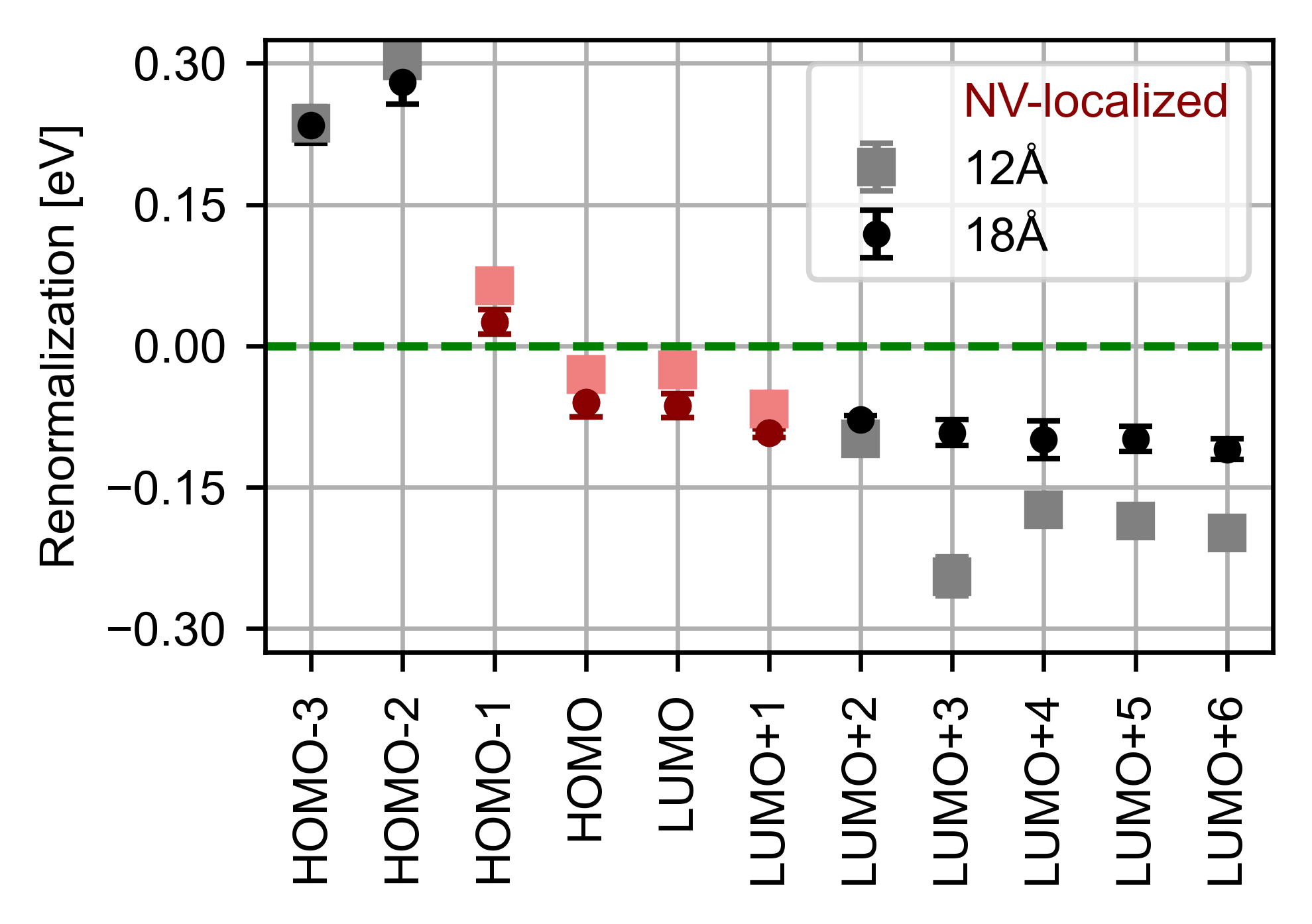}
\caption{Value of phonon renormalizations of the various single-particle energy levels in the $\beta$ spin channel for the F-terminated nanodiamond crystals (C\textsubscript{67}H\textsubscript{45}F\textsubscript{17}N) with 12 and 18 {\AA}  cell parameters as obtained from Monte Carlo simulations at 0 K using the PBE functional. Localization of states is defined as in Fig. \ref{fig:F_term_B3LYP}.}
\label{fig:phonon_renorm}
\end{figure}

{\it Quantum vibronic effects} -- Finally, we turn to the phonon renormalizations of the single-particle levels in a representative nanodiamond array: F-terminated nanodiamond (C\textsubscript{67}H\textsubscript{45}F\textsubscript{17}N), computed using a stochastic Monte Carlo (MC) sampling method.\cite{Kundu_JCTC_2023, Kundu_NV_JPCL_2024} Previously, we demonstrated that the PBE functional is qualitatively accurate in predicting the phonon renormalizations of various energy levels within NV$^-$ centers in bulk diamonds. \cite{Kundu_NV_JPCL_2024} Due to the computational expense of MC sampling, we only used the PBE functional for arrays of USNDs with cell lengths of 12 and 18 {\AA}, considering two different temperatures: 0 K and 300 K. Fig. \ref{fig:phonon_renorm} compares the renormalizations of single-particle  HOMO and LUMO energies for  F-terminated nanodiamonds at 0 K with cell lengths of 12 and 18 {\AA}.

We find that the renormalization of the valence band maximum and conduction band minimum (corresponding to HOMO-2 and LUMO +2 in our notation) of the array of nanodiamonds are about 300 meV and -100 meV, respectively,  resulting in a -400 meV renormalization of their gap. The magnitude of these renormalizations is similar to that found for the molecular crystal of a smaller diamondoid, pentamantane.\cite{Kundu_JCTC_2023}.

For the NV-localized levels within the band gap of the nanocrystal, we observe renormalizations with smaller magnitudes compared to the valence (HOMO-2 or below) and conduction (LUMO+2 or above) bands. In the case of a diamond crystal hosting an NV center, we also found that the magnitudes of the renormalization of the NV-localized defect levels are smaller than those of the valence and conduction bands.\cite{Kundu_NV_JPCL_2024} However, due to the dynamic Jahn-Teller (DJT) splitting of unoccupied defect levels, the latter exhibit more negative renormalizations compared to occupied orbitals, resulting in $\sim$ -150 meV renormalizations of the HOMO-LUMO gap ($\beta$-spin channel) and the related ($^3A_2 \rightarrow ^3E$) excitation energy.\cite{Kundu_NV_JPCL_2024} In contrast, the absence of degenerate levels in an array of nanodiamond clusters prevents DJT splitting from occurring. As a result, the occupied and unoccupied NV-localized levels exhibit similar and smaller renormalizations compared to those in the NV$^-$ -center in a diamond crystal, leading to an almost complete cancellation of quantum vibronic (phonon) effects on the gap or excitation energy. Interestingly, the small negative phonon renormalizations of unoccupied NV-localized levels lower their position relative to the hopping barrier,  by about  0.1 eV at 18 {\AA} distance, contributing to the charge stability of the point defect. However, for short distances between the nanoclusters, this stabilization effect is further reduced. 

%\section{Summary and Conclusions}\label{sec:conclusions}
In summary, we investigated the stability of the negatively charged nitrogen-vacancy center in ultrasmall nanodiamonds. Our findings indicate that the negatively charged nitrogen-vacancy center is unstable in hydrogen-terminated isolated nanodiamonds. These centers are likely to convert to a neutral charge state upon optical excitation due to the negative electron affinity of the system, consistent with what was previously observed in shallow nitrogen-vacancy centers on diamond surfaces.\cite{Kaviani_Nano_Lett_2014} Surface chemical modification, such as introducing electron-withdrawing groups through inductive effects, including fluorine, hydroxyl, or ether groups, can lead to a positive electron affinity and thus in principle be conducive to stabilizing the negative charge on the NV center. However, some of these groups exhibit electron withdrawing(donating) resonance (mesomeric) effects and lead to the introduction of acceptor states within the band gap of the nanoparticle; the presence of such states causes the charge to be transferred from the NV to the surface,  eventually quenching the fluorescence, despite the positive electron affinity of the nanoparticle. Since the presence of fluorine reduces spin coherence time of the spin defect significantly, for quantum information sensing applications, we propose a combination of hydroxyl and ether surface functionalization as a more effective approach to stabilize the negative charge on the nitrogen-vacancy center.

We find that the stability of the NV- is maintained for ultrasmall nanodiamonds arranged in arrays if the distance between the nanodiamonds is larger than their diameter. For shorter distances, charge and spin-hopping between nanodiamonds may occur, negatively impacting any spin readout process. Nevertheless, such hopping processes could be exploited for charge and spin-transport applications, including optical spin-valves.\cite{optical_spin_valve} Our calculations of the quantum vibronic coupling in USND arrays reveal that the valence and conduction bands undergo renormalizations similar to those observed in diamond or diamondoid crystals, resulting in a band gap reduction of around 400 meV. Due to the lack of symmetry in the array of nanodiamonds studied here, no degenerate defect levels are present and no dynamic Jahn-Teller splitting occurs. Consequently, the defect levels exhibit reduced phonon renormalizations compared to those observed for nitrogen-vacancy centers in the bulk diamond,\cite{Kundu_NV_JPCL_2024} leading to a minimal impact of quantum vibronic effects on the vertical excitation energy of the first triplet excited state. However, quantum vibronic effects
help stabilize the negative charge on individual nanoparticles within an array by lowering the energies of the lowest unoccupied molecular orbitals of the nanoparticles, thereby increasing the charge-hopping barrier. This stabilizing effect diminishes as the distances between nanodiamonds become comparable to their diameter.  Work is in progress to investigate the spin-coherence times of nanodiamond arrays.

\section{Methods}\label{sec:methods}
We performed all spin-unrestricted DFT calculations using the Qbox code,\cite{Qbox_Gygi_2008} employing norm-conserving pseudopotentials\cite{ONCV_2015} and a kinetic energy cutoff of 65 Ry. For the isolated nanoclusters, we consistently used a cubic cell with a length of 60 Bohr. To assemble nanoclusters into arrays, we utilized the open-source Python crystal builder PyXtal\cite{pyxtal} and visualized the structures with VESTA.\cite{vesta} We optimized all geometries using the Perdew-Burke-Ernzerhof (PBE) functional,\cite{PBE_Perdew_PRL_1996_1, PBE_Perdew_PRL_1996_2} applying a force tolerance of at least$10^{-3}$ Ha/Bohr for isolated nanoclusters and $10^{-4}$ Ha/Bohr for their molecular crystals. To compute the dynamical matrix, we employed the finite-difference method as implemented in the PyEPFD package,\cite{Kundu_PRM_2021, Kundu_JCTC_2023, pyepfd} using symmetric Cartesian displacements of $\pm 0.02$ and $\pm 0.04$ Bohr with the PBE functional. Based on these dynamical matrices, we generated canonical ensembles of displaced configurations at a specified temperature using Monte Carlo (MC) sampling,\cite{Kundu_JCTC_2023, Kundu_NV_JPCL_2024} as implemented in PyEPFD. We generated 600 independent samples and their antithetic pairs, yielding a total of 1200 configurations for each ensemble. We then performed single-point DFT/PBE calculations for each configuration using Qbox, averaging the single-particle energy levels across these configurations to obtain electron-phonon renormalization values.

\begin{acknowledgement}
This work was supported by MICCoM, as part of the Computational Materials Sciences Program funded by the U.S. Department of Energy, Office of Science, Basic Energy Sciences, Materials Sciences, and Engineering Division through Argonne National Laboratory, under Contract No. DE-AC02-06CH11357. This research used the resources of the University of Chicago Research Computing Center.
\end{acknowledgement}

\bibliography{references}
\end{document}